%% file: main.tex
\documentclass[letterpaper, 11pt]{article}
\usepackage[utf8]{inputenc}

\usepackage{geometry}
\usepackage[utf8]{inputenc}
\usepackage{amsmath,amssymb}
\usepackage{amsthm}
\usepackage{amssymb,graphicx,verbatim,enumitem,mdframed,mathrsfs}
\usepackage{mdframed}
\usepackage{subcaption}
\usepackage[toc]{appendix}

\usepackage{wrapfig}

\usepackage{algorithm}
\usepackage{algorithmic}
\usepackage{arydshln}

\newtheorem{definition}{\textbf{Definition}}
\newtheorem*{theorem*}{Theorem}

\providecommand{\mnorm}[1]{\ensuremath{\left\lvert#1\right\rvert}}

\def\eff{\mathsf{comEff}}
\def\pr{\mathsf{probF}}
\def\R{\mathbb{R}}
\def\Z{\mathcal{Z}}

\title{Randomized Reactive Redundancy for \\Byzantine Fault-Tolerance in Parallelized Learning}
\author{Nirupam Gupta and Nitin H. Vaidya}
\date{}

\begin{document}

\maketitle

\section{Introduction}
\begin{wrapfigure}{r}{0.45\textwidth}
  \begin{center}
    \includegraphics[width=0.4\textwidth]{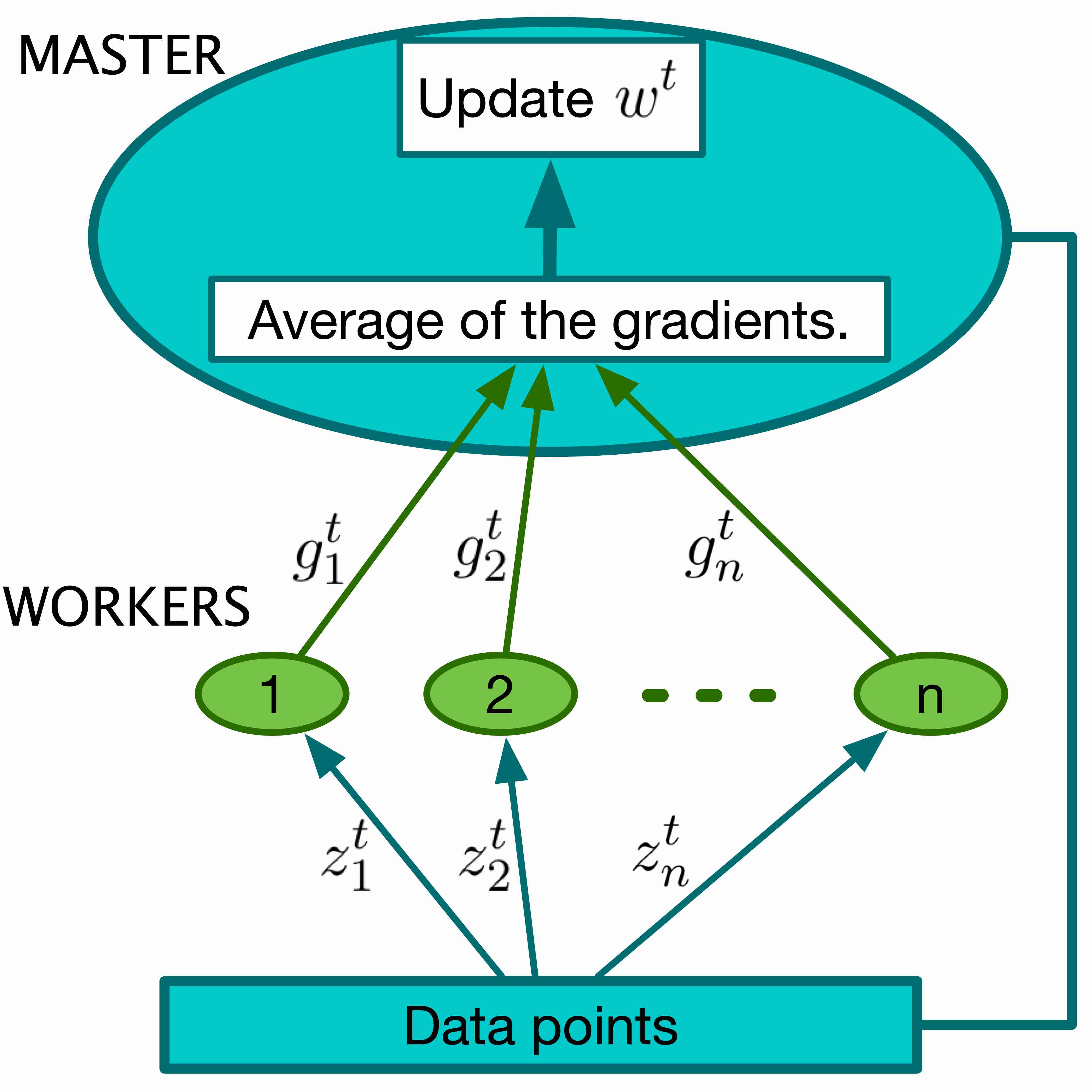}
  \end{center}
  \caption{\footnotesize{\it Parallelized-SGD method, for $m = n$, and $c_i(g^t_i) = g^t_i$ for all $i$.}}
  \label{fig:sys_arch}
\end{wrapfigure}

We consider the problem of Byzantine fault-tolerance in synchronous parallelized learning that is founded on the parallelized stochastic gradient descent (parallelized-SGD) method. \\

The system comprises a master, $n$ workers, and $N$ ($\gg n$) data points denoted by a set $\Z$. The system architecture is shown in Figure~\ref{fig:sys_arch}. Let $d$ be a positive integer, and let $\R^d$ denote the set of $d$-dimensional real-valued vectors. For a global parameter $w \in \R^d$, each data point $z \in \Z$ has a non-negative loss function $\ell(w, z) \in \R_{\geq 0}$. The goal for the master is to {\em learn} a parameter $w^*$ that is a minimum point\footnote{A local minimum point if the average loss function is non-convex, or a global minimum point if the average loss function is convex.} of the average loss evaluated for the data points. Formally, $w^*$ minimizes
\begin{align*}
    \frac{1}{N} \sum_{z \in \Z} \ell(w, z) ~
\end{align*}
in a {\em neighbourhood} of $w^*$. Although $w^*$ may not be the only minimum point, for simplicity $w^*$ denotes a minimum point for the average loss throughout this report.\\

The optimization framework forms the basis for most contemporary learning methods, including neural networks and support vector machines~\cite{bottou2018optimization}. 

\subsection{Overview of the parallelized-SGD method}
\label{sec:p_sgd}

{Parallelized-SGD} method is an expedited variant of the stochastic gradient descent method, an iterative learning algorithm~\cite{zinkevich2010parallelized}. In each iteration $t \geq 0$, the master maintains an estimate $w^t$ of $w^*$, and updates it using gradients of the loss functions for a certain number of randomly chosen data points at $w = w^t$. The details of the algorithm are as follows.

In each iteration $t$, the master randomly chooses a set of $m$ data points, denoted by $\Z_t \subset \Z$, and assigns $m_i$ data points to $i$-th worker for $i = 1, \ldots, \, n$, such that $\sum_{i = 1}^n m_i = m$. Let the data points assigned to the $i$-th worker in $t$-th iteration be denoted by $\{z^t_{i_1}, \ldots, \, z^t_{i_{m_i}}\}$. Each worker $i$ computes the gradients for the loss functions of its assigned points at $w^t$, 
\[g^t_{i_j} = \nabla \ell(w, z^t_{i_j}) \, |_{\,w = w^t}, ~ j = 1, \ldots, \, m_i ~, \]
and sends a {\em symbol} $c_i$, which is a function of its computed gradients $\{g^t_{i_1}, \ldots, \, g^t_{i_{m_i}}\}$, to the master. The master obtains the average value of the gradients for all the $m$ data points in $\Z_t$, 
\[g^t = \frac{1}{m}\sum_{z \in \Z_t}\nabla \ell(w, z) \, |_{\,w = w^t} ~ ,\]
as a function of the symbols $\{c_1, \ldots, \, c_n\}$ received from the workers. For example, if each worker $i$ sends symbol
\[c_i\left(g^t_{i_1}, \ldots, \, g^t_{i_{m_i}}\right) = \frac{1}{m_i} \sum_{j = 1}^{m_i} g^t_{i_j} ~ ,\]
then
\[g^t = \frac{1}{m}\sum_{i = 1}^n m_i c_i = \frac{1}{m}\sum_{z \in \Z_t}\nabla \ell(w, z) \, |_{\,w = w^t} ~ .\]
Upon obtaining $g^t$, the master updates the parameter estimate $w^t$ as 
\begin{align}
    w^{t+1} = w^t - \eta_t \left( \frac{1}{N}\sum_{i = 1}^N g^t_i \right), \label{eqn:update}
\end{align}
where $\eta_t$ is a positive real value commonly referred as the `step-size'. An illustration of the parallelized-SGD method is presented in Figure~\ref{fig:sys_arch} for the case when $m_i = 1, \, \forall i$.

\subsection{Vulnerability against Byzantine workers}

The above parallelized-SGD method is not robust against {\bf Byzantine} faulty workers. Byzantine workers need \underline{not} follow the master's instructions correctly, and might send malicious incorrect (or {\em faulty}) {symbols}. The identity of the Byzantine workers remains fixed throughout the learning algorithm, and is unknown {\em a priori} to the master. \\

We consider a case where up to $f$ ($< n/2$) of the workers are {Byzantine} faulty. {\bf Our objective} is design a parallelized-SGD method that has {\em exact fault-tolerance}, which is defined as follows.

\begin{definition}
A {parallelized-SGD} method has {\bf {exact} fault-tolerance} if the Master asymptotically converges to a minimum point $w^*$ exactly, despite the presence of Byzantine workers.
\end{definition}

\section{Proposed Solutions and Contributions}
We propose two coding schemes, one of which is {\bf deterministic} and the other is {\bf randomized}, for guaranteeing exact fault-tolerance if $2f < n$. Obviously, the master {\em cannot} tolerate more than or equal to $n/2$ Byzantine workers~\cite{chen2018draco}. Overviews of each these schemes are presented below. Before we proceed with the summary of our contribution and overviews of proposed coding schemes, let us define the {\em computation efficiency} of a coding scheme.

\begin{definition}
The {\bf computation efficiency} of a coding scheme is the {\em ratio of the number of gradients used for parameter update, given in~\eqref{eqn:update}, to the number of gradients computed by the workers in total}.
\end{definition}

For example, in each iteration of the parallelized-SGD method presented above, the total number of gradients computed by the workers is equal to $m$, and the master uses the average of all the $m$ gradients to update the parameter estimate~\eqref{eqn:update}. Therefore, the computation efficiency of a coding scheme (used for computing the symbols $c_1, \ldots, \, c_n$) in the traditional parallelized-SGD method is equal to $1$. \\

\noindent \fbox{
\parbox{0.97\textwidth}{
\subsection*{Summary of contributions:}

\begin{itemize}
    \item The {computation efficiency} of our {\bf deterministic coding scheme} is {twice as high} as that of a fault-correction code based scheme proposed by Chen et al., 2018~\cite{chen2018draco}. To improve upon the computation efficiency of the deterministic coding scheme, we propose a randomization technique.
    
    \item The computation efficiency of the {\bf randomized scheme} is {\em optimal} in {expectation}, and compares favorably to any coding scheme for tolerating Byzantine workers in the considered parallelized learning setting.
\end{itemize}
}}
~\\

\begin{figure}[htb!]
    \centering
    \includegraphics[width= 0.7 \textwidth]{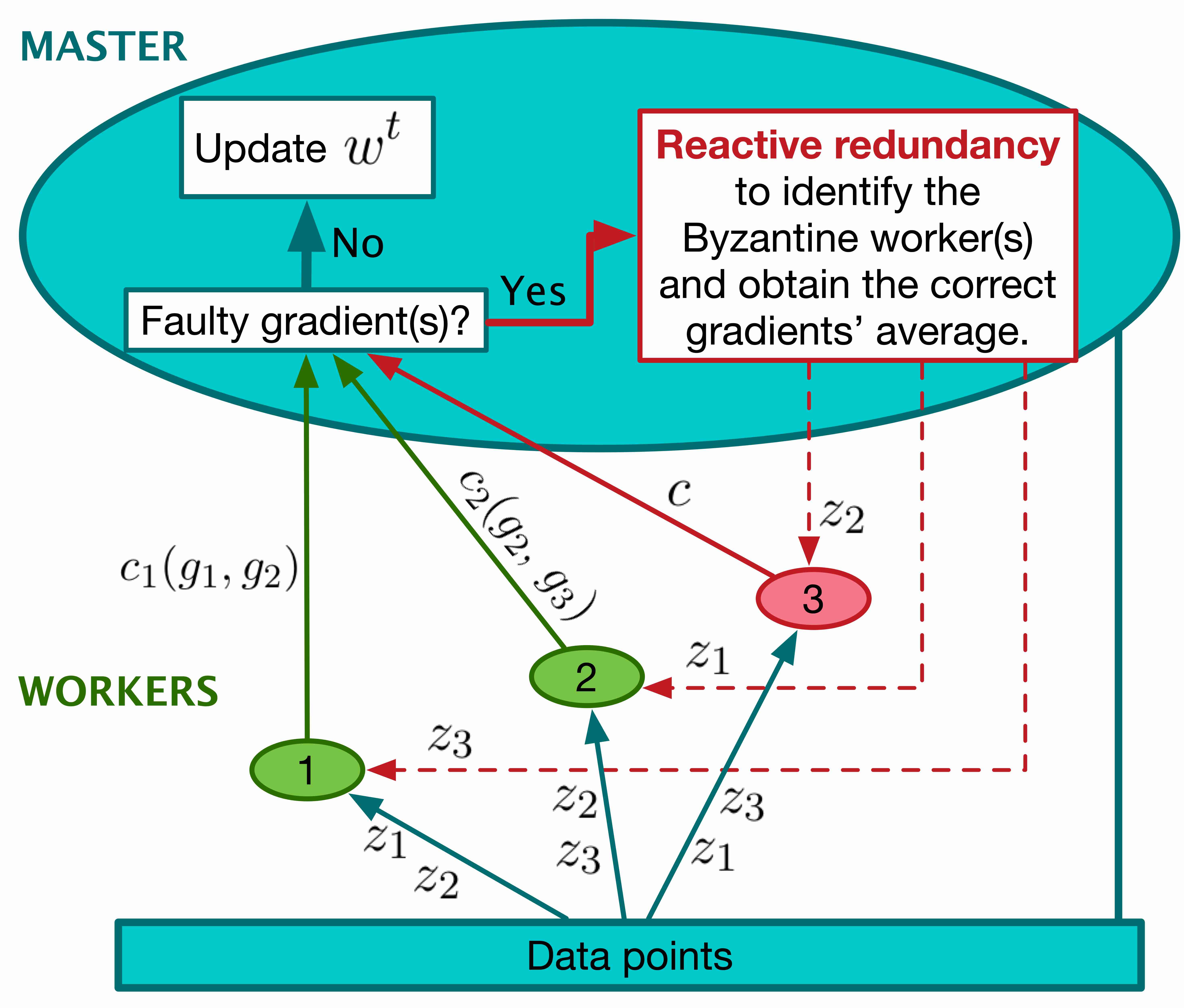}
    \caption{{\footnotesize{\it An {\bf example} of the deterministic coding scheme. Here, $n = 3$, and $f = 1$. For iteration $t$, the master assigns workers $1$, $2$ and $3$ data points $(z_1, ~ z_2)$, $(z_2, ~ z_3)$ and $(z_3, ~ z_1)$, respectively. Let, $g_1$, $g_2$ and $g_3$ denote the gradients for data points $z_1$, $z_2$ and $z_3$, respectively. The workers $1$, $2$ and $3$ are supposed to send {\em symbols} $c_1(g_1, g_2) = g_1 + 2g_2$, $c_2(g_2, g_3) = -g_2 + g_3$ and $c_3(g_3, g_1) = -g_1 - 2 g_3$, respectively. Let us ignore the arguments of $c_i$'s for the rest of the discussion. Note that $c_1 + c_2 = -(c_2 + c_3) = (1/2)(c_1 - c_3) = \sum_{i}g_i$. Therefore, the master can detect if a worker sends an faulty {\em symbol}. For instance, suppose that worker $3$ is Byzantine. If worker $3$ sends a symbol $c \neq c_3$ then $-(c_2 + c)$ and $(1/2)(c_1 - c)$ cannot be equal to $\sum_{i}g_i$ \underline{\em simultaneously}. This allows the master to detect \underline{if any} of the received symbols is faulty. However, mere fault-detection is not sufficient for identifying the Byzantine worker. For doing so, the master imposes a {\em reactive} redundancy in which each data point is assigned to an additional worker. Then, workers $1$, $2$, and $3$ are instructed to send symbols $u_1 = (c_2, \, c_3)$, $u_2 = (c_3, \, c_1)$, and $u_3 = (c_1, \, c_2)$. This enables to identify the Byzantine nature of worker $3$ (using majority voting on the symbols received), and consequentially recover the correct correct average of the gradients.}}}
    \label{fig:coding_scheme}
\end{figure}


\subsection{Overview of the deterministic scheme}
\label{sub:det}
For each iteration $t$, after choosing the data points, the master assigns each data point to $f+1$ workers. Each worker $i$ computes gradients for all its data points, and sends a symbol $c_i$ to the master such that, the collection of symbols $\{c_1, \ldots, c_n\}$ forms an $f$ {\em fault-detection} code, i.e. the master can detect up to $f$ faulty symbols, and the average of the gradients (for all the data points) is a function of the non-faulty symbols. Upon detecting any fault(s), the master imposes {\em reactive redundancy} where each data point (or data point specific to the detected fault(s)) is assigned to additional $f$ workers. Each worker now computes gradients for the additional data points assigned, and send symbols $u_1, \ldots, \, u_n$ that enables the master to identify up to $f$ faulty symbols in $\{c_1, \ldots,c_n\}$. Upon identifying the Byzantine workers that sent faulty symbols, the master can recover the correct average of the gradients. Hence, the scheme guarantees {\em exact fault-tolerance}. \\

A simple example illustrating the scheme is presented in Figure~\ref{fig:coding_scheme}. A {\em replication code} for the generic case is presented in Section~\ref{sec:scheme}.\\ 

We note the following generalizations, and drawback of the scheme.

\begin{itemize}
    \item \textbf{Generalizations:} 
    
    \begin{itemize}
        \item The workers may send symbols that are function of {\em compressed} gradients, proposed for improved communication efficiency in the {\em non-Byzantine} case~\cite{aji2017sparse, bernstein2018compressed, singh2019sparq, tang2019doublesqueeze}, instead of the original gradients.
        
        \item In general, any suitable fault detection code may be used in this scheme, we use a replication code as an example. The choice of the code will have impact on the communication and computation efficiency of the scheme. However, a deterministic scheme, that obtains {\em exact} fault-tolerance, cannot have computation efficiency greater than $1/(f+1)$ in all iterations.
    \end{itemize} 
    
    \item \textbf{Drawback:} In the deterministic scheme, each gradient is computed by $f+1$ workers even when all the $f$ Byzantine workers send non-faulty (or {\em correct}) symbols. In other words,
\begin{align*}
    \textbf{computation efficiency} = \frac{\text{\# gradients used for update}}{\text{\# gradients computed in total}} = \frac{1}{f+1},
\end{align*}
even when all the workers send correct symbols. This unnecessary redundancy can be significantly reduced by using a randomized approach presented below.
\end{itemize}

\begin{figure}[htb!]
     \centering
    \includegraphics[width= 0.65\textwidth]{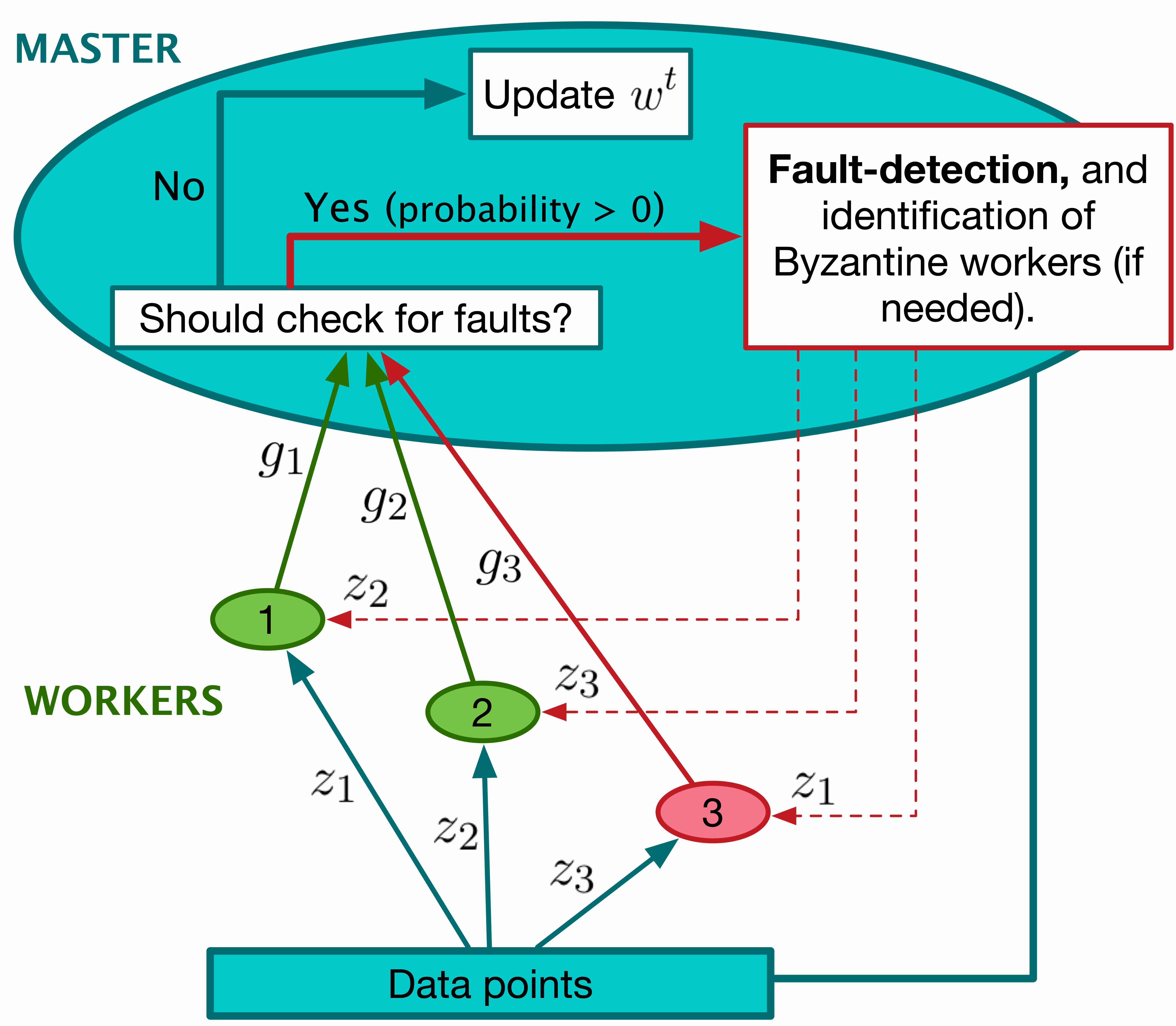}
    \caption{\footnotesize{An illustration of the randomized coding scheme. The workers $1$, $2$, and $3$ are supposed to send gradients $g_1$, $g_2$, and $g_3$, respectively, exactly as in the traditional parallelized-SGD method. Upon receiving the gradients, the master may check for faults with some positive probability less than $1$. For {\em fault-check}, each data point is assigned to an additional worker, and the honest workers follow the protocol of the deterministic scheme presented in Figure~\ref{fig:coding_scheme}.}}
    \label{fig:rand_code}
\end{figure}

\subsection{Overview of the randomized scheme} 
\label{sub:rand}

The master checks for faults only in intermittent iterations chosen at {\em random}, instead of all the iterations. Alternately, in each iteration, the master does a {\em fault-check} with some non-zero probability less than $1$. By doing so, the master significantly reduces the redundancy in gradients' computations whilst {\em almost surely} identifying the Byzantine workers that send faulty symbols {\em eventually}\footnote{As the parallelized-SGD method converges to the learning parameter regardless of the initial parameter estimate, a Byzantine worker that {\em eventually} stops sending faulty gradients poses no harm to the learning process. Hence, the master only needs to identify Byzantine workers that send faulty gradient(s) {\em eventually}.}. As in the deterministic scheme, upon detecting any fault(s) the master imposes reactive redundancy to identify the responsible Byzantine worker(s). However, correcting the detection fault(s) is optional. The identified Byzantine worker(s) are eliminated from the subsequent iterations. \\

An illustration of the scheme is presented in Figure~\ref{fig:rand_code}. Additional details for the generic case is presented in Section~\ref{sec:r_code}.\\

{\bf Significant savings on redundancy:} By reducing the probability of random fault-checks, the {\em expected} computation efficiency of the scheme can be made as close to $1$ as {\em desirable}. Note, a coding scheme that obtains exact fault-tolerance against a non-zero number of Byzantine workers \underline{\em cannot} have an expected computation efficiency of $1$.  \\

We note the following generalizations, and adaptation of the randomized scheme:
\begin{itemize}
    \item {\bf Generalization:} 
    \begin{itemize}
        \item Obviously, as in the deterministic case, the randomized scheme can be easily generalized for {\em compressed gradients}. 
        \item Instead of checking for faults for all the workers with equal probability, the master may use different probabilities for different workers. For doing so, workers can be assigned {\em reliability scores} as in the context of reliable crowdsourcing~\cite{raykar2012eliminating}. Other generalizations are presented in Section~\ref{sec:general}.
    \end{itemize}

    \item \textbf{Adaptation:} A lower probability of fault-checks implies higher probability of using faulty gradients for parameter update, and vice-versa. Higher probability of faulty updates means higher probability of slower convergence of the learning algorithm. To manage the trade-off between the {computation efficiency} and the rate of learning, we present an adaptive approach in Section~\ref{sec:adapt}. Essentially, the master may vary the probability of fault-checks -- depending upon the {\em observed} average loss at the current parameter estimate.
\end{itemize}

\section{Related works}

There has been some work on coding schemes for Byzantine fault-tolerance in parallelized machine learning, such as~\cite{chen2018draco, data2018data, rajput2019detox}. The scheme proposed by Data et al., 2018~\cite{data2018data}, however, is only applicable for loss functions whose arguments are linear in the learning parameter. The scheme, named DRACO, by Chen et al., 2018~\cite{chen2018draco} relies on {fault-correction codes} and so, has a computation efficiency of only $1/(2f + 1)$. At the expense of exact fault-tolerance, the computation efficiency of DRACO can be improved using gradient-filters~\cite{rajput2019detox}. Our randomized scheme has both; exact fault-tolerance, and favourable computation efficiency.\\

The fault-tolerance properties of the known gradient filters -- KRUM~\cite{blanchard2017machine}, trimmed-mean~\cite{yin2018byzantine}, median~\cite{yin2018byzantine}, geometric median of means~\cite{chen2017distributed}, norm clipping~\cite{gupta2019bft}, SEVER~\cite{diakonikolas2018sever}, or others~\cite{mhamdi2019fast, prasad2018robust} -- rely on additional assumptions either on the distribution of the data points or the fraction of Byzantine workers. Moreover, the existing gradient-filters {\em do not} obtain exact fault-tolerance {\em unless} there are redundant data points.\\

To the best of our knowledge, none of the prior works have proposed the idea of {\em reactive redundancy} for tolerating Byzantine workers efficiently in the context of parallelized learning. In other contexts, such as checkpointing and rollback recovery, mechanisms that combine proactive and reactive redundancy have been utilized. For instance, Pradhan and Vaidya \cite{DBLP:journals/tc/PradhanV97} propose a mechanism where a small number of replicas are utilized proactively to allow detection of faulty replicas; when a faulty replica is detected, additional replicas are employed to isolate the faulty replicas.


\input{reactive_code.tex}

\input{optimist_code.tex}

\section{Generalizations of the Randomized Coding Scheme}
\label{sec:general}

Our randomized scheme can be generalized as follows.

\begin{itemize}
    \item {\bf Variants of the parallelized-SGD method:} We can use the randomized scheme even for different variants of the parallelized-SGD method where workers send {\em compressed} or {\em communication-efficient} gradients, as proposed in~\cite{aji2017sparse, bernstein2018compressed, singh2019sparq, tang2019doublesqueeze}. 
    
    \item {\bf Self-checks:} Instead of imposing reactive redundancy, the master can compute the gradients on its own, and compare them with the gradients received from the workers to check for faults. Similarly as above, the master may optimize the additional workload by choosing the probability of fault-checks adaptively as presented in Section~\ref{sec:adapt}.
    
    \item {\bf Selective fault-checks:} Gradients (or symbols) that are outliers amongst the received gradients (or symbols) should be checked for faults with relatively higher probability. Additionally, the master can assign {\em reliability scores} to the workers, as done in the context of reliable crowdsourcing~\cite{raykar2012eliminating}. Symbols from workers with lower reliability scores should be checked for faults with higher probability. 
    
    \item {\bf Gradient-filters:} The master can further improve on the computation efficiency by combining the randomized coding scheme with lightweight gradient-filters~\cite{gupta2019byzantine, mhamdi2019fast, yin2018byzantine}. When using gradient-filters, the master does not have to identify all the Byzantine workers. This idea has been explored in Rajput et al., 2019~\cite{rajput2019detox} for a deterministic coding scheme.
    
    \item {\bf {\em Distributed} learning framework:} Our randomized scheme can also be used for Byzantine fault-tolerance in {\em distributed} learning framework, where the data points are {distributed} amongst the workers, i.e. two workers may have different sets of data points~\cite{chen2017distributed, yin2018byzantine}. In this case, besides checking for faulty gradient(s), the master must also {\em validate} the data points used by the workers for computing the gradients in the first place. As most existing data validation tools are computationally expensive~\cite{downs2010your, JMLR:v18:15-650, raykar2012eliminating, vuurens2011much}, the master may use our randomized scheme to optimize the trade-off between the cost of data validation and the convergence-rate of a distributed learning algorithm.
\end{itemize}

\section{Summary}
\label{sec:summ}
In this report, we have presented two coding schemes, a deterministic scheme and a randomized scheme, for exact Byzantine fault-tolerance in the parallelized-SGD learning algorithm. \\

In the deterministic scheme, the master uses a fault-detection code in each iteration. Upon detecting any fault(s), the master imposes reactive redundancy to correct the faults and identify the Byzantine worker(s) responsible for the fault(s). \\

The randomized scheme improves upon the computation efficiency of the deterministic scheme. Here, the master uses fault-detection codes only in randomly chosen {\em intermittent} iterations, instead of all the iterations. By doing so, the master is able to optimize the trade-off between the {\em expected} computation efficiency, and the convergence-rate of the parallelized learning algorithm. 


\section*{Acknowledgements}
Research reported in this paper was sponsored in part by the Army Research Laboratory under Cooperative Agreement W911NF- 17-2-0196, and by National Science Foundation award 1610543. The views and conclusions contained in this document are those of the authors and should not be interpreted as representing the official policies, either expressed or implied, of the the Army Research Laboratory, National Science Foundation or the U.S. Government.

\bibliographystyle{plain}
\bibliography{ref.bib}



\end{document}

%% file: reactive_code.tex
\section{Coding Schemes}
In this section, we present a specific deterministic scheme for the generic case, and present further details for the randomized scheme.

\subsection{Deterministic coding scheme}
\label{sec:scheme}
As an example of the deterministic scheme, we use a {\em replication code}. For simplicity, suppose that none of the Byzantine workers have been identified until iteration $t \geq 0$. Then, the scheme for the $t$-iteration is as follows.\\

The master (randomly) chooses $m$ data points, and assigns each data point to $f+1$ workers. Thus, each worker, on average, gets $m(f+1)/n$ data points. Upon computing the gradients for all its data points (at $w^t$), each worker sends a {\em symbol}; a tuple of its computed gradients. Consequentially, the master receives $f+1$ copies (or replicas) of each data point's loss function's gradient. As there are at most $f$ Byzantine workers, the master can detect if the received copies of a gradient are faulty by simply comparing them with each other. Suppose that the copies of the gradient of a particular data point $\hat{z}$ are faulty (i.e. they are not unanimous). Then, the master imposes {\em reactive redundancy} where it re-assigns $\hat{z}$ to $f$ additional workers that compute and send additional $f$ copies of the gradient for $\hat{z}$. Upon acquiring $2f+1$ copies of $\hat{z}$'s gradient, the master can not only obtain the correct gradient by majority voting, it also identify the responsible Byzantine worker(s). Ultimately, the master recovers the correct gradients for all the $m$ data points, and updates $w^t$ as~\eqref{eqn:update}.\\

The identified Byzantine worker(s) are eliminated from the subsequent iterations. Upon updating $f$ and $n$, the above scheme is repeated for the $(t+1)$-iteration.

\subsubsection*{Computation efficiency}


Let $\kappa_t$ be the number of Byzantine workers identified until the $t$-th iteration. If the master does not detect a fault in the $t$-th iteration then the computation efficiency of the scheme is $1/(f - \kappa_t + 1)$. Otherwise, the worst-case computation efficiency is $1/(2(f - \kappa_t) + 1)$. \\

As there are at most $f$ Byzantine workers, the master will detect faults and impose {\em reactive redundancy} in at most $f$ iterations. Thus, for $t > f$ iterations, the computation efficiency of the scheme is greater than or equal to $1/(f+1)$ for {\em at least} $t - f$ iterations. In case $T \gg f$, the average computation efficiency of the scheme is {\em effectively} greater than or equal to $1/(f+1)$. \\

\textbf{Note:} We would like to reiterate the fact that a deterministic coding scheme with {computation efficiency} greater than $1/(f+1)$, in all iterations, \underline{\em cannot} have exact fault-tolerance against at most $f$ Byzantine workers~\cite{lindell2010introduction}. However, communication efficiency can be improved using other codes.

%% file: optimist_code.tex
\subsection{Randomized coding scheme}
\label{sec:r_code}


In the randomized scheme, the master checks for faults (and does identification of Byzantine worker if needed) only for randomly chosen intermittent iterations. In each iteration, the master runs the traditional parallelized-SGD method by default. However, before updating the parameter estimate, the master decides to check for faults in the received symbols (or gradients) with probability $q > 0$. Fault-checks and identification of Byzantine workers (if needed) is done using the protocol outlined for the deterministic coding scheme in Section~\ref{sec:scheme}.\\


For the purpose of analysis, assume that each Byzantine worker $i$ tampers its gradient(s) independently in each iteration with probability at least $p_i > 0$. Then, $i$ remains unidentified by the master after $t$ iterations with probability less than or equal to $\left(1 - {q \, p_i} \right)^t$, 
which approaches $0$ as $t$ approaches $\infty$. In other words, $i$ gets identified almost surely. This holds for all Byzantine workers that tamper gradient(s) eventually.



\subsubsection*{Computation efficiency}

As the master checks for faults with probability $q>0$ in each iteration, the {\em expected} computation efficiency of the randomized scheme is greater than or equal to
\begin{align}
    (1 - q) \times 1 + q \times \frac{1}{2f+1} = 1 - q \left(\frac{2f}{2f+1}\right). \label{eqn:efficiency}
\end{align}
The above lower bound for the {\em expected} computation efficiency is computed by assuming the worst-case where the master imposes $2f$ redundancy for each gradient in the fault-detection phase. The actual computation efficiency will be larger than this lower bound. However, this lower bound suffices to understand the benefits of our coding scheme.\\

From above, the {\em expected} computational efficiency of the randomized coding scheme can be made as close to one as desirable by choosing $q$ appropriately. Specifically, for a $\delta > 0$, let
\[q = \delta\left(\frac{2f + 1}{2f}\right) \leq 1.\]
Then, the {\em expected} computational efficiency of the randomized coding scheme is greater than or equal to $1 - \delta$. 

\subsubsection*{Efficiency versus convergence-rate}
Smaller probability of fault-checks $q$ implies higher efficiency, as is evident from~\eqref{eqn:efficiency}. However, smaller $q$ also means higher probability of using faulty gradient(s) for updating the parameter estimate, which could result in slower convergence of the learning algorithm. \\

Suppose that each Byzantine worker chooses to tamper its gradient(s) independently with probability $p > 0$, then the probability of a faulty update in the $t$-th iteration (assuming none of the Byzantine workers have been identified yet) equals
\begin{align}
    &(\text{probability of faulty gradients}) \times (\text{probability of \underline{not} checking for fault(s)}) \nonumber \\ 
    &= \left(1 - (1 - p)^f \right) \times (1 - q) \label{eqn:prob_faults}
\end{align}
Therefore, determining an {\em optimal} value of $q$ is a multi-objective optimization problem where; 
\begin{itemize}
    \item {\bf Objective 1}: \underline{maximize the expected computation efficiency}, given by~\eqref{eqn:efficiency}.
    \item {\bf Objective 2}: \underline{minimize the probability of faulty updates}, given by~\eqref{eqn:prob_faults}.
\end{itemize}
~

Obviously, the above objectives cannot be met simultaneously. That is, there does not exist a $q$ that maximizes and minimizes the expected computation efficiency and the probability of faulty updates, respectively, at the same time. This trade-off between the computation efficiency and the reliability (or correctness) of the updates can be managed by the following adaptive approach.

\subsection{Adaptive randomized coding}
\label{sec:adapt}

Let $\eff_t(q)$ and $\pr_t(q)$ denote the {\em expected} computation efficiency and the probability of faulty update in iteration $t$, if the probability of doing a fault-check equals $q$. Let $\kappa_t$ denote the number of identified Byzantine workers until iteration $t$. By substituting $f$ by 
\[f_t = f - \kappa_t\] 
in~\eqref{eqn:efficiency} and~\eqref{eqn:prob_faults}, we obtain
\begin{align*}
    \eff_t(q) = \frac{2f_t(1-q) + 1}{2f_t +1}, \text{  and  } \pr_t(q) = \left(1 - (1 - p)^{f_t} \right) \times (1 - q)
\end{align*}
Note, maximizing $\eff_t(q)$ is equivalent to minimizing $(1 - \eff_t(q))^2$, and minimizing $\pr_t(q)$ is equivalent to minimizing $(\pr_t(q))^2$. Thus, the probability of fault-check in the $t$-iteration, denoted by $q^*_t$, is given by the minimum point of the weighted average of $(1 - \eff_t(q))^2$ and $(\pr_t(q))^2$, i.e.,  
\begin{align}
    q^*_t = \arg \min_{q \in [0, \, 1]} (1 -  \lambda_t) \, \left(1 - \eff_t(q) \right)^2 + \lambda_t \, \left(\pr_t(q) \right)^2 ~, \label{eqn:obj}
\end{align}
where $\lambda_t \in [0, \,1]$. Higher value of $\lambda_t$ (greater than $1/2$) implies that minimizing $\pr_t(q)$ takes precedence over maximising $\eff_t(q)$, and vice versa. 

\subsubsection*{Choice of $\lambda_t$}
We note that a suitable value of $\lambda_t$ can be computed using the average loss, denoted by $\ell_t$, computed over the chosen data points at the current parameter estimate. Specifically, if $\Z_t$ denotes the set of data points chosen and $w^t$ denotes the current parameter estimate in the $t$-iteration, then
\[\ell_t = \frac{1}{\mnorm{\Z_t}}\sum_{z \in \Z_t} \ell(w^t, \, z).\]
Then,
\begin{align}
    \lambda_t = ( 1- e^{-\ell_t}).  \label{eqn:lamdba}
\end{align}
If $\lambda_t$ is given by~\eqref{eqn:lamdba}, then for higher {\em observed} loss $\ell_t$ minimizing the probability of faulty updates takes precedence. This is quite intuitive as the master would prefer the updates to fault-free when the observed loss is high, for improved convergence-rate to the learning parameter. \\

The following boundary conditions further justify the choice of $\lambda_t$ given by~\eqref{eqn:lamdba}.



\begin{itemize}
    \item As $\ell_{t}$ approaches $\infty$, $\lambda_t$ approaches $1$. In this extreme case, 
    \[q^{*}_t = \arg \min_{q \in [0, \, 1]} \left( \pr_t(q) \right)^2 = 1\]
    Thus, the master checks for faults in {\em almost} all iterations when the observed loss $\ell_t$ is extremely high.
    
    \item If $p = 0$, i.e. Byzantine workers do not tamper their gradients with certainty, 
    \[q^{*}_t = \arg \min_{q \in [0, \, 1]} \left( \eff_t(q) \right)^2 = 0.\]
    Obviously, if the gradients received from the Byzantine workers are correct with {\em certainty} then there is no need for fault-checks. Similarly, if $\kappa_t = f$, i.e. the master has identified all the $f$ Byzantine workers, then 
    \[q^{*}_t = \arg \min_{q \in [0, \, 1]} \left( \eff_t(q) \right)^2 = 0.\]

\end{itemize}

{\bf Note:} For saving on the computation cost, the master may use the workers for computing $\ell_t$ in parallel. However, in this case  the master would only be able to obtain an approximation of $\ell_t$, instead of the actual value, as up to $f$ of the workers are Byzantine. Nevertheless, approximate $\ell_t$ suffices for the above adaptation. An approximation of $\ell_t$ can be computed by taking the truncated or trimmed mean of the average loss evaluated by the workers for their respective data points~\cite{wilcox2011introduction}. 



%% file: main.bbl
\begin{thebibliography}{10}

\bibitem{aji2017sparse}
Alham~Fikri Aji and Kenneth Heafield.
\newblock Sparse communication for distributed gradient descent.
\newblock {\em arXiv preprint arXiv:1704.05021}, 2017.

\bibitem{bernstein2018compressed}
Jeremy Bernstein, Yu-Xiang Wang, Kamyar Azizzadenesheli, and Anima Anandkumar.
\newblock signsgd: Compressed optimisation for non-convex problems.
\newblock {\em arXiv preprint arXiv:1802.04434}, 2018.

\bibitem{blanchard2017machine}
Peva Blanchard, Rachid Guerraoui, Julien Stainer, et~al.
\newblock Machine learning with adversaries: {Byzantine} tolerant gradient
  descent.
\newblock In {\em Advances in Neural Information Processing Systems}, pages
  119--129, 2017.

\bibitem{bottou2018optimization}
L{\'e}on Bottou, Frank~E Curtis, and Jorge Nocedal.
\newblock Optimization methods for large-scale machine learning.
\newblock {\em Siam Review}, 60(2):223--311, 2018.

\bibitem{chen2018draco}
Lingjiao Chen, Hongyi Wang, Zachary Charles, and Dimitris Papailiopoulos.
\newblock {DRACO}: Byzantine-resilient distributed training via redundant
  gradients.
\newblock In {\em International Conference on Machine Learning}, pages
  903--912, 2018.

\bibitem{chen2017distributed}
Yudong Chen, Lili Su, and Jiaming Xu.
\newblock Distributed statistical machine learning in adversarial settings:
  {Byzantine} gradient descent.
\newblock {\em Proceedings of the ACM on Measurement and Analysis of Computing
  Systems}, 1(2):44, 2017.

\bibitem{data2018data}
Deepesh Data, Linqi Song, and Suhas Diggavi.
\newblock Data encoding for {Byzantine}-resilient distributed gradient descent.
\newblock In {\em 2018 56th Annual Allerton Conference on Communication,
  Control, and Computing (Allerton)}, pages 863--870. IEEE, 2018.

\bibitem{diakonikolas2018sever}
Ilias Diakonikolas, Gautam Kamath, Daniel~M Kane, Jerry Li, Jacob Steinhardt,
  and Alistair Stewart.
\newblock {SEVER}: A robust meta-algorithm for stochastic optimization.
\newblock {\em arXiv preprint arXiv:1803.02815}, 2018.

\bibitem{downs2010your}
Julie~S Downs, Mandy~B Holbrook, Steve Sheng, and Lorrie~Faith Cranor.
\newblock Are your participants gaming the system?: screening mechanical turk
  workers.
\newblock In {\em Proceedings of the SIGCHI conference on human factors in
  computing systems}, pages 2399--2402. ACM, 2010.

\bibitem{gupta2019byzantine}
Nirupam Gupta and Nitin~H Vaidya.
\newblock Byzantine fault-tolerant distributed linear regression.
\newblock {\em arXiv preprint arXiv:1903.08752}, 2019.

\bibitem{gupta2019bft}
Nirupam Gupta and Nitin~H Vaidya.
\newblock Byzantine fault-tolerant parallelized stochastic gradient descent for
  linear regression.
\newblock {\em 57th Annual Allerton Conference on Communication, Control, and
  Computing}, 2019.

\bibitem{JMLR:v18:15-650}
Srikanth Jagabathula, Lakshminarayanan Subramanian, and Ashwin Venkataraman.
\newblock Identifying unreliable and adversarial workers in crowdsourced
  labeling tasks.
\newblock {\em Journal of Machine Learning Research}, 18(93):1--67, 2017.

\bibitem{lindell2010introduction}
Yehuda Lindell.
\newblock Introduction to coding theory lecture notes.
\newblock {\em Department of Computer Science Bar-Ilan University, Israel
  January}, 25, 2010.

\bibitem{mhamdi2019fast}
El~Mahdi~El Mhamdi, Rachid Guerraoui, and Arsany Guirguis.
\newblock Fast machine learning with byzantine workers and servers.
\newblock {\em arXiv preprint arXiv:1911.07537}, 2019.

\bibitem{DBLP:journals/tc/PradhanV97}
Dhiraj~K. Pradhan and Nitin~H. Vaidya.
\newblock Roll-forward and rollback recovery: Performance-reliability
  trade-off.
\newblock {\em {IEEE} Trans. Computers}, 46(3):372--378, 1997.

\bibitem{prasad2018robust}
Adarsh Prasad, Arun~Sai Suggala, Sivaraman Balakrishnan, and Pradeep Ravikumar.
\newblock Robust estimation via robust gradient estimation.
\newblock {\em arXiv preprint arXiv:1802.06485}, 2018.

\bibitem{rajput2019detox}
Shashank Rajput, Hongyi Wang, Zachary Charles, and Dimitris Papailiopoulos.
\newblock {DETOX}: A redundancy-based framework for faster and more robust
  gradient aggregation.
\newblock In {\em Advances in Neural Information Processing Systems}, pages
  10320--10330, 2019.

\bibitem{raykar2012eliminating}
Vikas~C Raykar and Shipeng Yu.
\newblock Eliminating spammers and ranking annotators for crowdsourced labeling
  tasks.
\newblock {\em Journal of Machine Learning Research}, 13(Feb):491--518, 2012.

\bibitem{singh2019sparq}
Navjot Singh, Deepesh Data, Jemin George, and Suhas Diggavi.
\newblock {SPARQ-SGD}: Event-triggered and compressed communication in
  decentralized stochastic optimization.
\newblock {\em arXiv preprint arXiv:1910.14280}, 2019.

\bibitem{tang2019doublesqueeze}
Hanlin Tang, Xiangru Lian, Tong Zhang, and Ji~Liu.
\newblock Doublesqueeze: Parallel stochastic gradient descent with double-pass
  error-compensated compression.
\newblock {\em arXiv preprint arXiv:1905.05957}, 2019.

\bibitem{vuurens2011much}
Jeroen Vuurens, Arjen~P de~Vries, and Carsten Eickhoff.
\newblock How much spam can you take? an analysis of crowdsourcing results to
  increase accuracy.
\newblock In {\em Proc. ACM SIGIR Workshop on Crowdsourcing for Information
  Retrieval (CIR’11)}, pages 21--26, 2011.

\bibitem{wilcox2011introduction}
Rand~R Wilcox.
\newblock {\em Introduction to robust estimation and hypothesis testing}.
\newblock Academic press, 2011.

\bibitem{yin2018byzantine}
Dong Yin, Yudong Chen, Kannan Ramchandran, and Peter Bartlett.
\newblock Byzantine-robust distributed learning: Towards optimal statistical
  rates.
\newblock In {\em International Conference on Machine Learning}, pages
  5636--5645, 2018.

\bibitem{zinkevich2010parallelized}
Martin Zinkevich, Markus Weimer, Lihong Li, and Alex~J Smola.
\newblock Parallelized stochastic gradient descent.
\newblock In {\em Advances in neural information processing systems}, pages
  2595--2603, 2010.

\end{thebibliography}
